# Charting an embryological path to cancer cure: A discussion of disease hallmarks

Jaime Cofre*

Laboratory of Molecular Embryology and Cancer, Federal University of Santa Catarina, room 313b, Florianópolis, SC, 88040-900, Brazil

* Corresponding author. Laboratory of Molecular Embryology and Cancer, Universidade Federal de Santa Catarina, sala 313b, Florianópolis, SC, 88040-900, Brazil

E-mail address: jaime.cofre@ufsc.br

## Abstract

Embryology has long played a foundational role in shaping our scientific understanding of animal evolution. In recent decades, growing evidence has also highlighted its role in cancer. Despite the indisputable similarities between embryonic development and cancer, there has been limited discussion on the profound embryological implications for the disease. This article explores the understanding of cancer as an embryological and evolutionary phenomenon, offering a fresh perspective on the disease and discussing immediate consequences in the search for therapeutic approaches

## 1. Introduction

The vast majority of research and articles on cancer genetics are overwhelmingly consistent about the role of mutations in disease onset and progression [1]. However, the traditional model of tumorigenesis reduces cancer to a proliferative phenomenon [2,3], overlooking substantial evidence that a multidisciplinary approach is needed to understand the disease [4]. The hallmarks of cancer have been identified, and their mutational genetic underpinnings are clearly defined (Figure 1a) [2]. Some of the most neglected aspects of cancer are related to embryology and animal evolution [5]. The next section will discuss how embryology and animal evolution can illuminate the paths in the search for therapeutic alternatives for cancer patients.

## 2. The embryological hallmarks of cancer

Embryo formation is prompted by a biomechanical phenomenon of cell fusion (fertilization), allowing cells to proliferate while physically joined by surface molecules (e.g., cadherins) until the formation of the early-stage morula and blastocyst. The embryo continues its development, performing collective migration movements during epiboly (second stage of embryonic development). Gastrulation begins via epithelial–mesenchymal transition (EMT) (third stage, morphogenesis), which represents a critical point of convergence (common ground) for clinical pathologists, oncologists, and embryologists. EMT reveals the striking similarity between human gastrulation and metastasis, given that both processes begin with the breakdown of the basal lamina and are triggered by mechanical forces and physical phenomena [5]. The embryo is capable of differentiating into all human cell types, including germ cells, which contain the reproductive program needed to reproduce the embryonic process in the next generation. As discussed below, cancer also recapitulates key embryonic stages, including fertilization. Its most intriguing features, such as the Warburg effect and extracellular acidification, may find explanations in animal embryogenesis and its deep evolutionary history.

In line with the concept of embryogenesis recapitulation, many cancers are associated with cell fusion molecules and appear to mimic aspects of fertilization [6]. The process of cell fusion dramatically accentuates oncogenic characteristics [7,8]. For example, high metastatic capacity is observed in hybrids formed by the fusion of non-metastatic tumor cells with bone marrow-derived macrophages [9], as well as from the fusion of normal human gastric mucosal cells with umbilical cord mesenchymal stem cells separated from umbilical cord blood [10]. In the latter case, hybridization induced oncogenic transformation, leading to a marked increase in proliferative capacity, migration, and invasive potential compared with parental cells [10]. Thus, fusibility is not limited to tumor cells, and the in vitro or in vivo context is not determinant [11]. The fact that tumor cells fuse spontaneously represents an enigmatic phenomenon still awaiting an explanation from the scientific community [12] , despite being recognized for over a century [13].

Continuing with the idea of the importance of embryogenesis in cancer, it is imperative to consider that cancer cells also recapitulate characteristics found in embryonic stem cells (ESCs), which are present during early human development as part of the blastocyst epiblast [14]. Tumors found to be poorly differentiated by histological analysis overexpress genes commonly enriched in ESCs, as well as underexpress genes regulated by Polycomb, an arguably embryonic gene. Additionally, poorly differentiated tumors more frequently overexpress activation targets of Nanog, Oct4, Sox2, and c-Myc than well-differentiated tumors. For instance, this expression profile has been identified in breast cancer, poorly differentiated glioblastomas and bladder carcinomas [14]. These findings demonstrate an unprecedented link between genes associated with ESC identity and tumor histopathology, further supporting the concept of embryogenesis recapitulation in cancer. On the other hand, Myc is among the most infamous cancer oncogenes. Its deregulation may be related to most, if not all, human cancers [15]. A distinctive feature of Myc is that its dysregulation typically does not involve mutations. Interestingly, its participation in the Warburg effect in cancer may represent an evolutionary feature of embryonic genes, as will be discussed later.

An argument supporting the recapitulation of embryonic epiboly is the recognition of several modes of cancer cell migration, such as amoeboid, filopodial, mesenchymal, and collective amoeboid movements [16]. It could not be otherwise, given that cancer, according to my proposal, is quintessentially an embryonic phenomenon. Understandably, this line of reasoning gave rise to the notion that embryonic developmental processes are recapitulated in most types of epithelial and mesenchymal cancers [17]. In my view, cancers as diseases recapitulate their own embryonic origin.

The connections between EMT, cancer, and embryo development [18–20] are indisputable and well-grounded in scientific research, potentially explaining why developmental biologists commonly pursue research into oncology. Clinical oncologists have pointed out alleged contradictions in the classical EMT model as a universal mechanism in tumor invasion and metastasis, such as incomplete EMT and reversion to an epithelial phenotype. These apparent contradictions arise only within human embryology and can be elucidated through the broader perspective of animal embryology. Incomplete EMT occurs during the embryonic development of Nematostella vectensis [21]. Reversion to an epithelial phenotype is observed in the mesenchymal–epithelial transition (MET) that triggers the formation of embryonic structures [22]. Furthermore, collective migration of human cells finds support in the embryonic morphogenetic movements of various animal groups [23].

It is important to emphasize that adopting an evolutionary perspective can be useful for understanding how the disease cancer has acquired different properties throughout animal evolution, potentially contributing to the development of therapeutic solutions in human oncology. A practical example is that extracts from basal animals are being researched as potential sources of new antitumor drugs [24]. The most promising agent currently under development for the treatment of lung cancer (Manzamine A) is derived from the marine sponge Haliclona sp [25]. In my view, new anticancer drugs should be analyzed in the context of their role within the embryo (not just as an extract), as it is widely agreed that embryos themselves have the capacity to inhibit metastasis [26,27]. Also, in accordance with this evolutionary perspective, in oncological research, it was demonstrated for the first time that amoeboid movement, characteristic of our protist origin, is not a form of migration but rather a cellular state [28,29]. Presently, mesenchymal–amoeboid [30], epithelial–amoeboid [31], and epithelial–mesenchymal–amoeboid transitions [32] are recognized in some cancers. These transitions are consistent with the evolutionary origin of cancer. Epithelial–mesenchymal–amoeboid transitions, in particular, evoke the behavior of embryo cells that carry [33] and modify [34] the extracellular matrix while migrating.

Delving further into the intersection between cancer and embryology, one of the most striking aspects is the incorporation of a reproductive program to promote oncogenesis (soma-to-germline transition) [35]. Scientific research on oncology indicates that tumorigenesis involves a soma-to-germline transition, which may contribute to the acquisition of neoplastic characteristics [36,37]. Several studies on different animal models, such as Drosophila melanogaster [38], mouse [39], and humans [40], showed that tumors exhibit characteristics similar to those of germline cells. This phenomenon only makes sense with the understanding of cancer as an embryological phenomenon, much overlooked in oncological research. The only biological system in which the soma-to-germline transition is indispensable is the animal embryo, which forms germ cells capable of reconstructing embryogenesis in the next generation. Without an embryological perspective, our understanding of cancer remains limited; however, without an evolutionary perspective, the search for effective cures becomes even more challenging. For example, germline genes that drive oncogenesis in D. melanogaster (fruit fly) [38] are associated with more clinically aggressive tumors in humans [40]. Additionally, cancer was triggered by the activation of some human orthologs from germline genes of D. melanogaster [38]. These findings demonstrate the close relationship between cancer and animal evolution. It is now well established that the soma-to-germline transition is a common hallmark of human cancer [37] and, by all indications, a conserved feature across all metazoans (Figure 1b) [5].

Finally, two critical questions remain: How can the Warburg effect in cancer be explained from an embryological perspective? And how can the acidification of the extracellular environment by cancer cells be explained? Copepods, representatives of marine zooplankton, were the first animals whose high resistance to hypoxia was associated with the Warburg effect [41]. Therefore, their metabolism can be considered similar to tumor systems. For example, Tigriopus californicus, an inhabitant of supralittoral rock pools exposed to anoxic conditions (~0 mg/L) [42], has become a physiological model for the study of oxygen deprivation in animals that lack respiratory structures and have secondarily lost key members of the hypoxia-inducible factor (HIF) pathway [43]. Studies on T. californicus demonstrated a shift from oxidative phosphorylation to glycolysis under hypoxic conditions and a significant increase in mRNA levels of pyruvate dehydrogenase kinase (PDK) [43]. In other words, the Warburg effect (Warburg 1956) occurs naturally in animals living in highly hypoxic environments. This phenomenon, in my opinion, reveals the ancestral conditions of animal genesis and places the Warburg effect within a clearly evolutionary context.

Despite the large volume of articles and hypotheses aimed at understanding the Warburg effect in cancer, its function remains unclear [44], and the perspective of cancer as an evolutionary phenomenon is completely disregarded in oncology. The Warburg effect may have enabled extensive cell proliferation and embryonic growth under the anoxic conditions of ancient oceans, which directly contradicts the idea that oxygenation was an environmental barrier to animal evolution [45,46]. Surprisingly, among genes strongly upregulated in response to hypoxia, Pdk1 is a potential target of Myc [47]. Myc is one of the genes involved in early embryogenesis [14] and an embryonic signature of oncogenesis. Given that Myc activates glycolytic genes, including lactate dehydrogenase, it is believed to participate in the Warburg effect [48]. This is consistent with the proposal that the Warburg effect would support high and efficient cell proliferation for embryo formation under the hypoxic conditions of early animal evolution [49]. The answer to the acidification of the extracellular medium of cancer cells may also lie within an evolutionary context. An alkaline marine environment has been proposed as the most plausible setting for the origin of animal life [50], likely characterized by active proton efflux mechanisms. Voltage-gated proton channels (Hv1) with ΔpH-dependent regulation are present in a broad range of organisms, including protists and humans [51]. These channels may have contributed to calcification [52,53] and acidification of ancient oceans. It is known that Hv1 channels drive malignant progression in diverse cancers [54]. Inhibition of proton efflux by Hv1 channels was found to decrease cell proliferation and invasiveness, as well as impair the ability to acidify the extracellular medium.

Many of the mysteries of cancer seem to be intertwined with our evolutionary past. For evolution enthusiasts, cancer uncovers fundamental cellular processes that reflect our ancestral origins. Cancer seems to have been shaped throughout evolutionary history in an embryonic context. Only by considering these new and challenging scenarios can innovations emerge for the definitive resolution of the disease and its cure.

## 3. Concluding remarks

Cancer is a major cause of mortality worldwide, affecting both rich or developing countries [55,56]. One of the most critical challenges for patients is the reduction in life expectancy, particularly due to metastasis, which remains a major cause of cancer-related deaths [56]. Despite significant advances over the past decade, the complex and heterogeneous nature of cancer continues to hinder the development of effective therapies and therapeutic combinations capable of producing lasting responses in the majority of patients [57]. This last and important statement represents a long-standing observation: "The dilemma of cancer research is exemplified by the increasing obscurity of much of the writing, by the extraordinary remoteness, range, and intricacy of the lists of papers presented at cancer meetings and by their failure to illuminate the scene. Information accumulates space while understanding lags behind" [58].

Thus, information based on genetic reductionism has been accumulated over 60 years, to no avail. Future advances in cancer therapy will depend, in part, on acknowledging that cancer is not merely an individual disease, nor one that can be effectively addressed by targeting isolated molecular components. Rather, this disease has acquired its characteristics throughout animal phylogenetic history and must, therefore, be understood as an inherently evolutionary phenomenon. Therapeutic advances will only be meaningful if we accept its embryological nature, exploring innovative solutions in the great animal transitions. Such approaches may ultimately bring relief to patients diagnosed with cancer. While the genetic and molecular characteristics of cancer are important, they fall within a broader cellular and developmental context that is rooted in embryology. Technically, embryonic cells do not acquire genetic mutations during embryogenesis but undergo "controlled metastasis" shaped by the embryonic environment. Introducing cancer cells into blastocysts reverses their malignant phenotype [26,27], which underscores the potential of animal embryogenesis in guiding future therapeutic strategies.

Figures

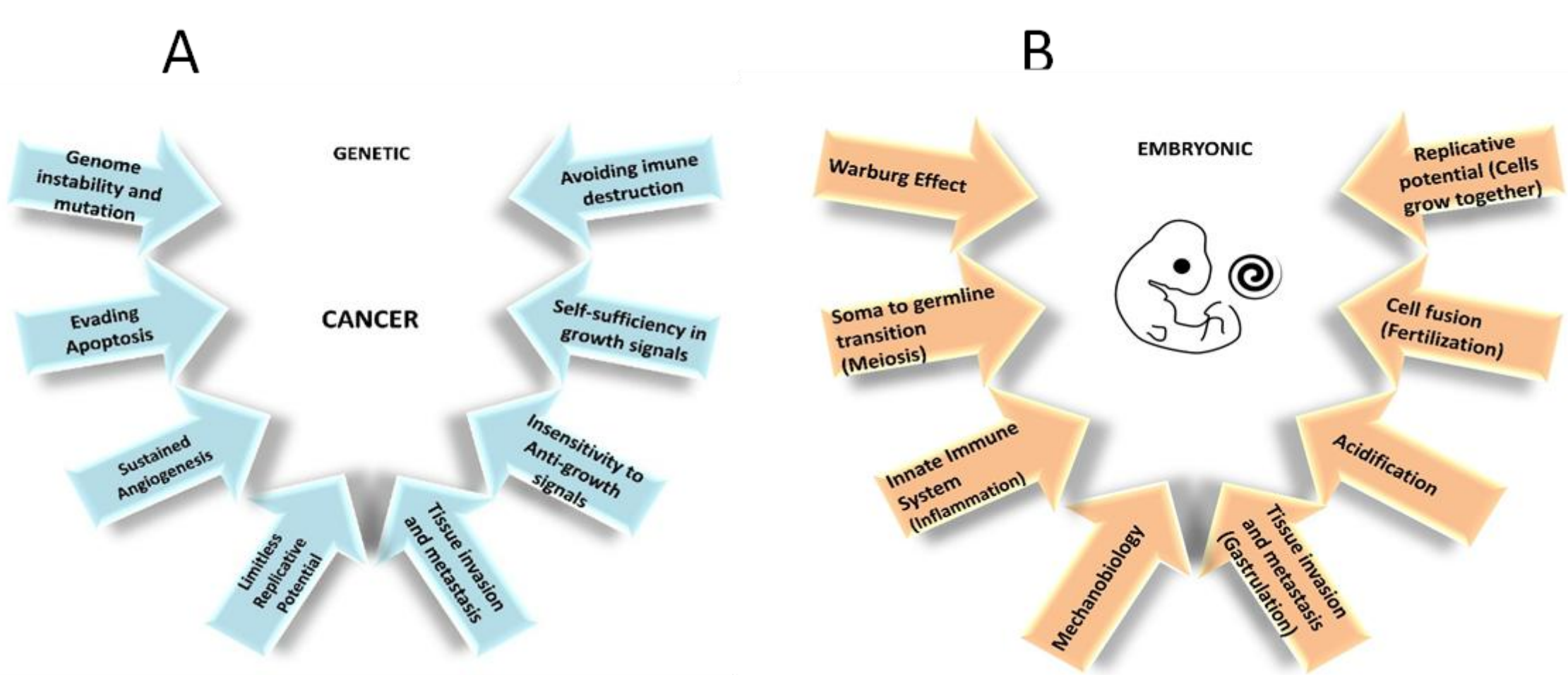

Figure 1. Hallmarks of cancer. (a) This panel presents the traditional hallmarks of cancer, with a strong emphasis on a genetic-based explanatory mechanism. (b) This panel also considers the embryonic hallmarks of cancer. The embryo shown in the image is only an example of an animal embryo. The symbol at the center represents the structural coherence first established in the early embryo and preserved throughout animal evolution. All cancer hallmarks are required for embryo formation. Some reflect the direct influence of the environment on animal evolution, such as the Warburg effect and acidification. These environmental conditions are defined as embryological hallmarks, consistent with the author's central hypothesis that views the embryo as a benign tumor. Cell fusion under hypoxic conditions can induce both multiflagellate fusion and polyspermy. Likewise, cell fusion under hypoxic conditions reveals the importance of syncytia in animal evolution. The incorporation of a reproductive program is reflected in the transition from the somatic lineage to the germline, where meiosis occurs. Replicative potential is demonstrated both in the context of cells growing together as well as in the well-known collective movements of cancer cells. Epithelial–mesenchymal transition links metastasis and gastrulation as closely related phenomena. The influence of the innate immune system, mechanical processes, and biophysical aspects is depicted in the figure but is not developed in the present article.